\documentclass{ws-p9-75x6-50}

\usepackage{graphicx}
\begin{document}
\title{Relativistic Thomas-Fermi Model at Finite Temperatures}
\author{Gianfranco Bertone}

\address{I.C.R.A. International Center for Relativistic Astrophysics,
I-00185 Rome, Italy \\ Institut d'Astrophysique, F-75014 Paris, France
\\
 Department of Astrophysics, University of Oxford, NAPL, Oxford OX13RH,
UK \\
E-mail: bertone@astro.ox.ac.uk}

\author{Remo Ruffini}
\address{I.C.R.A. International Center for Relativistic Astrophysics and
Physics Department, University ``La Sapienza'' , I-00185 Rome, Italy \\
E-mail: ruffini@icra.it}
\maketitle

\abstracts{
We briefly review the Thomas-Fermi statistical model of atoms
in the classical non-relativistic formulation and in the generalised
finite-nucleus relativistic formulation. We then discuss the classical 
generalisation of the model to finite temperatures in the non-relativistic 
approximation and present a new relativistic model at finite temperatures,
investigating how to recover the existing theory in the limit
of low temperatures. This work is intended to be a propedeutical
study for the evaluation of equilibrium configurations of
relativistic ``hot'' white dwarfs.}

\section{Introduction}

The Thomas-Fermi statistical model of the atom  has been extensively
used, since its early formulation in 1927 \cite{fermi,thomas},
to evaluate the equation of state of compressed matter and in general
in the description of the electronic structure of atoms and solids
under different physical conditions \cite{lund}. In our previous work
\cite{berto} we applied a relativistic generalisation of this model
\cite{ruffini,ferre} to the evaluation of the equation of state and
the equilibrium configurations of cold white dwarfs, whereas ``cold'' 
means that the temperature
of the star is much smaller than the Fermi energy of the degenerate
electron gas. We are now interested in the finite temperature effects
on the equation of state and we thus derive a temperature dependent
formulation of the relativistic Thomas-Fermi model.
The paper is organised as follows: first (section 2) we briefly describe
the
classical Thomas-Fermi model, i.e. the non-relativistic
and completely degenerate case. Then (section 3) we'll review the finite
temperature
formulation of this problem , giving the corresponding exact equation
and its approximated version. We'll thus pass to the relativistic
theory in the completely degenerate case
(section 4) and finally to the new relativistic model at finite 
temperature (section 5).

\section{Classical Thomas-Fermi model}

We first consider the simple Thomas-Fermi model. Let us consider the
spherically symmetric problem of a nucleus with Z protons and A nucleons
interacting with a fully degenerate gas of Z electrons.

The fundamental equation of electrostatics for this problem is
\begin{equation}
\Delta V(r)=4 \pi e n_e(r)
\label{elettro}
\end{equation}
where $V(r)$ is the electrostatic potential and  $n_e(r)$ is the number density of electrons.

The electrostatic potentialis related to the Fermi momentum (and thus to the number density of electrons) by  the equilibrium condition (see ~\cite{lali})
\begin{equation}
p_{F}^{2}(r)/2m-eV(r) = \mbox{const} \equiv E_F
\label{equili}
\end{equation}
where the name $E_F$ stands for \emph{Fermi-Thomas chemical potential}
or \emph{Fermi Energy} of the electrons.

To put the equation \ref{elettro} in adimensional form we introduce the
new function $\Phi(r)$, related to the coulomb potential by
\begin{equation}
\Phi(r)=V(r)+E_F/e
\label{4}
\end{equation}
and the corresponding adimensional function $\chi$, implicitly defined
by
\begin{equation}
\Phi(r)=\frac{Ze\chi}{r}
\label{6}
\end{equation}

Furthermore we introduce the new independent variable x, related to the
radius r by the relation  $r=bx$, where
\begin{equation}
b=(3\pi)^{2/3}\frac{\hbar^2}{me^2}\frac{1}{2^{7/3}}\frac{1}{Z^{1/3}}
\label{eq:6}
\end{equation}

It is easy to show that in this case one can write eq.\ref{elettro} in
the form
\begin{equation}
\frac{d^2 \chi}{dx^2}=\frac{\chi^{3/2}}{x^{1/2}}
\label{thomfer}
\end{equation}
which is the classical adimensional form of the Thomas-Fermi equation.

The first initial condition for this equation follows from the request
that approaching the nucleus one gets the ordinary Coulomb potential
\begin{equation}
\chi(0)=1
\label{con}
\end{equation}

The second condition comes from the normalisation condition
\begin{equation}
N=\int_0^{r_0} 4 \pi n_e r^2 dr
\end{equation}
which gives
\begin{equation}
N=Z\left[x_0\chi'(x_0)-\chi(x_0)+1 \right]
\label{coni}
\end{equation}
and for neutral atoms ($N=Z$)
\begin{equation}
x_0\chi'(x_0)=\chi(x_0)
\label{cono}
\end{equation}

\section{Temperature Dependent Non-relativistic Model}

We introduce now the finite temperature effects in the model. This was
already done in 1940 by Marshack and Bethe \cite{marsha} through a
perturbation treatment, while the full adimensional equation is
discussed in a successive work of Feynman, Metropolis and Teller
\cite{feyn}.

The density of an electron gas at temperature T can be written as
\begin{equation}
n_e=\frac{\sqrt{2}m_e^{3/2}}{\pi^2 \hbar^3} \int_0^\infty
\frac{\sqrt{\epsilon}}{e^\frac{\epsilon-\mu}{kT} +1} d \epsilon =
\frac{\sqrt{2}m_e^{3/2}}{\pi^2 \hbar^3} (KT)^{3/2} I_1\left(\frac{\mu}{kT}
\right)
\label{nonrelT}
\end{equation}
where
\begin{equation}
I_1(x) = \int_0^\infty \frac{\sqrt{y}}{e^{y-x} +1} dy
\end{equation}

Using the same adimensional variables introduced in the previous section
and introducing the temperature parameter $\tau$ as
\begin{equation}
\tau = \frac{b}{Ze^2} KT
\end{equation}
we can rewrite the electron density in adimensional form
\begin{equation}
n_e=\frac{\sqrt{2}m_e^{3/2} Z^{3/2}e^3 \tau^{3/2}}{\pi^2 \hbar^3
b^{3/2}} I_1\left(\frac{\chi}{\tau x} \right)
\end{equation}

We can thus express the electrostatic equation in the following
adimensional form
\begin{equation}
\frac{d^2 \chi}{dx^2}=\frac{3}{2} \tau^{3/2} x  I_1\left(\frac{\chi}{\tau
x} \right)
\end{equation}

This equation is formally different from the one obtained in the case of
complete degeneracy, nevertheless it can be easily shown that if one
develops the integral which appears in eq. \ref{nonrelT} for small
temperatures (see appendix for details) one gets the following formula
at the first order
\begin{equation}
\frac{d^2 \chi}{dx^2}=\frac{\chi^{3/2}}{x^{1/2}} \left[ 1+
\frac{\pi^2}{8}\frac{\tau^2 x^2}{\chi^2} + ... \right]
\end{equation}
where we neglect terms of the order $O(\tau^4)$.

\section{Relativistic model at T=0}

When considering a relativistic extension of the model the finite size
of the nucleus must be taken in account to avoid the central singularity
(see e.g. \cite{berto}). In this case we have thus to introduce a new
term into the fundamental equation of electrostatics, representing the
positive distribution of charges in the interior of the nucleus $n_p(r)$

\begin{equation}
\Delta V(r)=4 \pi e n_e(r)-4 \pi e n_p(r)
\label{elettro2}
\end{equation}

The quantities V(r) and n(r) are in this case related by
\begin{equation}
c\sqrt{p_{F}^{2}+m^2c^2}-eV(r) = \mbox{const} \equiv E_F
\label{equili2}
\end{equation}

Using eq.\ref{4} it is possible to put eq. \ref{equili2} in the form
\begin{equation}
p^2_F=\frac{e^2}{b} \Phi^2+ 2me \Phi
\end{equation}
which, using \ref{6},becomes
\begin{equation}
p_F=2mc\left(\frac{Z}{Z_{cr}}\right)^{2/3}\left(\frac{\chi}{x}\right)^{1/2}\left[1+
\left(\frac{Z}{Z_{cr}}\right)^{4/3}\frac{\chi}{x} \right]^{1/2}
\label{impu}
\end{equation}
where
\begin{equation}
Z_{cr}=\left(\frac{3\pi}{4}\right)^{1/2}\left(\frac{\hbar
c}{e^2}\right)^{3/2}\approx 2462.4
\end{equation}

Remembering the relation between the Fermi momentum and the number
density of a fermion gas
\begin{equation}
n_e=\frac{p_{F}^{3}}{3\pi^2\hbar^3}
\label{2}
\end{equation}
we obtain the following expression
\begin{equation}
n_e=\frac{Z}{4\pi b^3}\left(\frac{\chi}{x}\right)^{3/2}\left[1+
\left(\frac{Z}{Z_{cr}}\right)^{4/3}\frac{\chi}{x} \right]^{3/2}
\label{adens}
\end{equation}

We can also express the second term of the right-hand side of
eq.\ref{elettro} in terms of adimensional quantities: we assume here an
homogeneous spherical nucleus, with a radius given by the approximate
formula
\begin{equation}
r_{nuc}=1.2A^{1/3} \: fm
\end{equation}

The number density of protons is therefore
\begin{equation}
n_{p}=\frac{3Z}{4\pi r_{nuc}^3}\Theta\left(x_{nuc}-x \right)
\end{equation}

Finally we can write eq.\ref{elettro2} in the form
\begin{equation}
\frac{d^2 \chi}{dx^2}=\frac{\chi^{3/2}}{x^{1/2}} \left[1+
\left(\frac{Z}{Z_{cr}}\right)^{4/3}\frac{\chi}{x}
\right]^{3/2}-\frac{3x}{{x_{nuc}}^3}\Theta\left(x_{nuc}-x \right)
\label{thomfer2}
\end{equation}
where $x_{nuc}$ is the adimensional size of the nucleus
($r_{nuc}=bx_{nuc}$).

Equation \ref{thomfer} is what we call \lq \emph{Generalised
adimensional Fermi-Thomas equation}\rq.

The first initial condition for this equation follows from the fact that
$ \chi \propto r \Phi $ and therefore $\chi
\stackrel{r{\rightarrow}0}{\longrightarrow}0$, and so
\begin{equation}
\chi(0)=0
\label{con1}
\end{equation}

The second condition comes from the normalisation condition
\begin{equation}
N=\int_0^{r_0} 4 \pi n_e r^2 dr= Z \int_0^{x_0}
\frac{\chi^{3/2}}{x^{1/2}} \left[1+
\left(\frac{Z}{Z_{cr}}\right)^{4/3}\frac{\chi}{x} \right]^{3/2}\;\; x
\;\; dx
\end{equation}
with $r_0=bx_0$ atom size. Developing this formula we have
\begin{equation}
N = Z \int_0^{x_{nuc}} x \chi'' \;\; dx + \frac{3Z}{x_{nuc}^3}
\int_0^{x_{nuc}} x^2 \;\; dx + Z \int_{x_{nuc}}^{x_0} x \chi'' \;\; dx
\end{equation}
which gives again the relation
\begin{equation}
N=Z\left[x_0\chi'(x_0)-\chi(x_0)+1 \right]
\label{con2}
\end{equation}

Note that the physical quantities such as the coulomb potential and the
density of electrons do not show any singularity in the center, neither
on the border of the nucleus, being dependent just on the function
$\chi$ and his first derivative. The only discontinuity appears in the
second derivative of $\chi$ due to our rough assumption of homogeneous
spherical nucleus.

It is also evident that the scaling properties of the classical
Thomas-Fermi equation are lost in this case, where one has to integrate
the adimensional equation separately for each  value of Z and different
values of $x_0$, i.e. for different states of compression.

In figg.1 and 2 we show an example of the applications of this model to
the study of equilibrium configurations of cold White Dwarfs. For more
details see ref. \cite{berto}.

\begin{figure}[t]
\resizebox{\hsize}{!}{\includegraphics{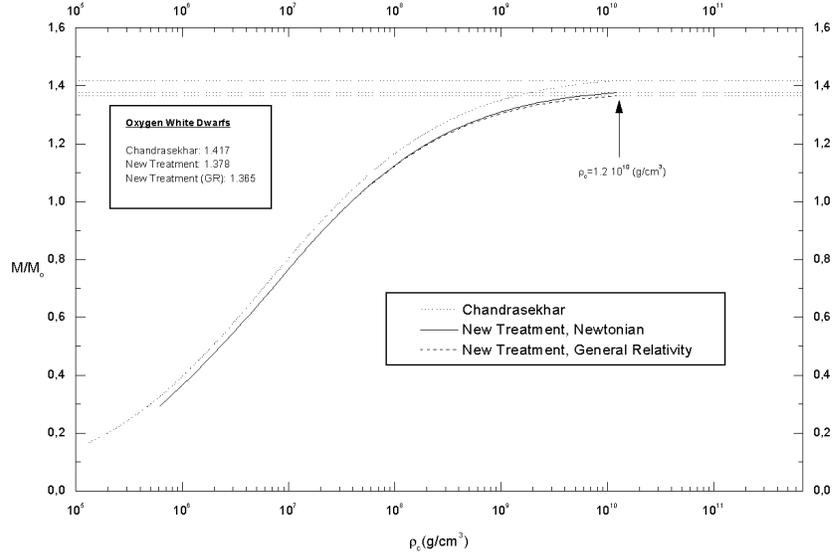}}
\caption[fig]{Equilibrium configurations curve for Oxygen WD in the
$M/M_{\odot} -  \rho_c$ plane, obtained in Newtonian theory and General
relativity. For comparison we show the results obtained with
Chandrasekhar model .Reproduced from Bertone \& Ruffini\cite{berto}.}
\label{ossigeno}
\end{figure}

\begin{figure}[t]
\resizebox{\hsize}{!}{\includegraphics{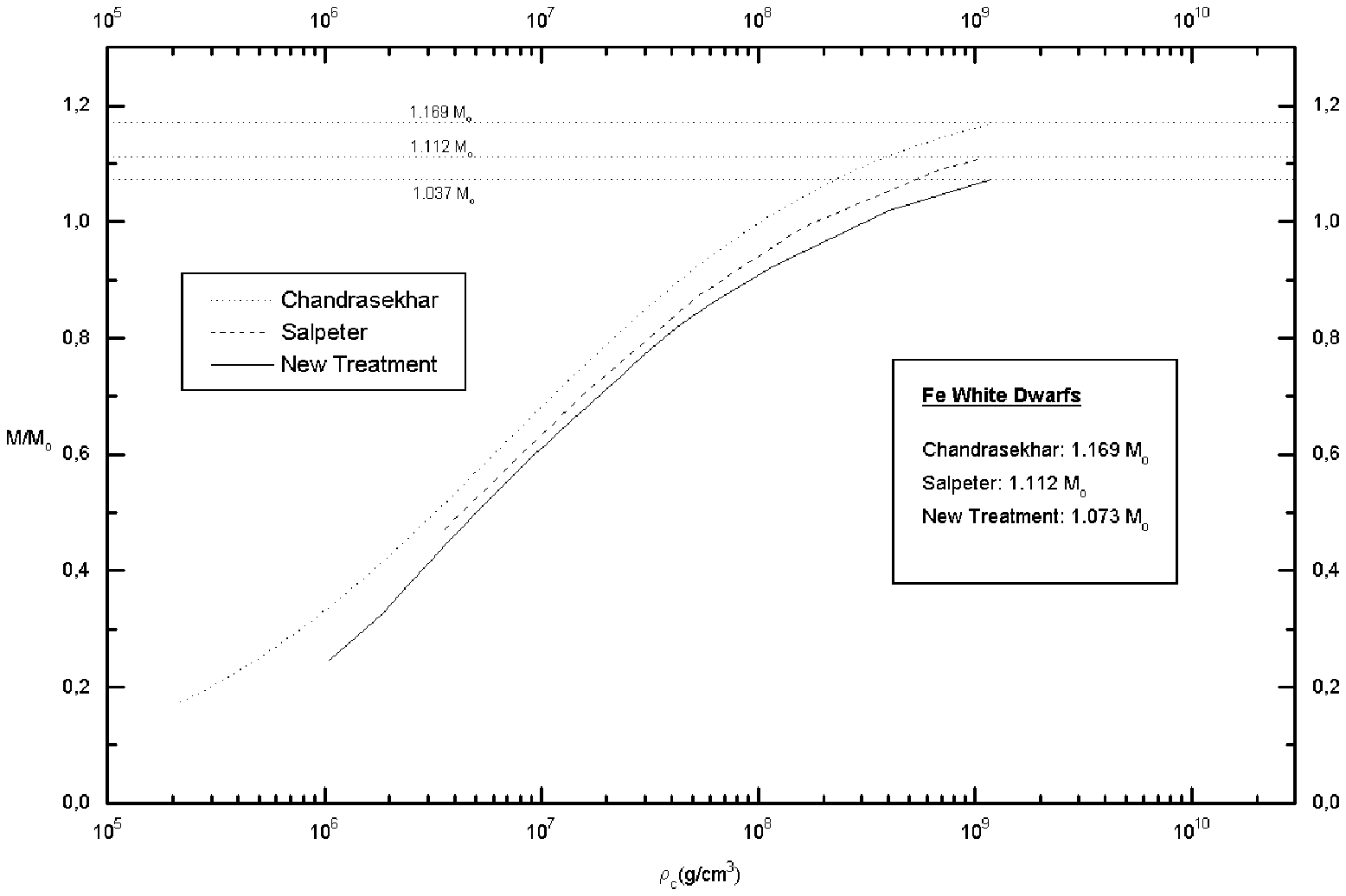}}
\caption[fig]{Equilibrium configurations curve for Iron WD in the
$M/M_{\odot} - \rho_c$ plane, obtained in Newtonian theory (the General
relativistic curve is pratically superposed at the newtonian one at
these low densities). For comparison we show the results obtained by
Chandrasekhar and Salpeter .Reproduced from Bertone \&
Ruffini\cite{berto}.}
\label{ferro}
\end{figure}

\section{Temperature Dependent Relativistic Model}

We consider now the complete problem of a relativistic and degenerate
gas of electrons at temperature T surrounding a positively charged
nucleus.

The number density of such a gas is
\begin{equation}
n_e=\frac{1}{\pi^2 (c \hbar)^3} \int_{m c^2}^\infty
\frac{\sqrt{\epsilon^2 - m^2 c^4}\epsilon}{e^\frac{\epsilon-\mu}{kT} +1}
d \epsilon = \frac{(KT)^3}{\pi^2 (c\hbar)^3} I_2\left(\frac{\mu}{KT}
\right)
\label{relT}
\end{equation}
where
\begin{equation}
I_2(x) = \int_a^\infty \frac{y \sqrt{y^2-a^2}}{e^{y-x} +1} dy
\end{equation}
and a is the fixed parameter $a=m c^2/(KT)$.

Introducing adimensional variables and the temperature parameter $\tau$
we can write now
\begin{equation}
n_e=\frac{1}{b^3}\frac{Z^3}{\pi^2}\left( \frac{e^2}{c \hbar}\right)^3
\tau^3 I_2\left(\frac{\chi}{\tau x} \right)
\end{equation}
and inserting the previously defined quantity $Z_{cr}$ we find
\begin{equation}
n_e=\frac{3Z}{4 \pi b^3} \left( \frac{Z}{Z_{cr}}\right)^2 \tau^3
I_2\left(\frac{\chi}{\tau x} \right)
\end{equation}

The final and more general expression of the Thomas-Fermi equation is
thus found to be
\begin{equation}
\frac{d^2 \chi}{dx^2}= 3 \tau^{3} x \left( \frac{Z}{Z_{cr}}\right)^2
I_2\left(\frac{\chi}{\tau x}
\right)-\frac{3x}{{x_{nuc}}^3}\Theta\left(x_{nuc}-x \right)
\end{equation}

This formula is the main result of this paper and will be the point of
departure for the evaluation of the equation of state of compressed
matter in extreme conditions of high temperatures in a relativistic
regime.

We will present elsewhere the numerical integration of this equation and

the corresponding equilibrium configurations of hot relativistic white
dwarfs.

\section{Conclusions}

We have first discussed the classical Thomas-Fermi method and the
corresponding generalisation to finite temperature $or$ to the
relativistic regime. A finite nucleus treatment including both effects
has then been presented. This work must be considered as
propedeutical to the evaluation of the equilibrium configurations of
relativistic hot white dwarfs. The numerical integration of the highly
non-linear equation obtained and the application to the equation of
state of white dwarfs matter will be presented elsewhere.

\section*{Appendix}

Following Landau and Lifshitz \cite{lali} we recall here how to
approximate for low temperatures integrals of the form
\begin{equation}
I= \int_0^\infty \frac{f(\epsilon)}{e^\frac{\epsilon-\mu}{KT} +1} d
\epsilon
\end{equation}
appearing in the statistical treatment of fermions.

Without going into details, we just recall that using the fact that
$\mu/T>>1$ one can put the former integral in the form
\begin{equation}
I=\int_0^\mu f(\epsilon) d \epsilon + 2(KT)^2 f'(\mu) \int_0^\infty
\frac{z}{e^z +1} dz + \frac{1}{3}(KT)^4 f'''(\mu) \int_0^\infty
\frac{z^3}{e^z +1} dz + ...
\end{equation}

These integrals can be evaluated as
\begin{equation}
\int_0^\infty \frac{z^{x-1}}{e^z +1} dz= (1 -2^{1-x})\Gamma(x)
\sum_{n=1}^\infty \frac{1}{n^x}
\end{equation}

We thus obtain our approximate result
\begin{equation}
I=\int_0^\mu f(\epsilon) d \epsilon + \frac{\pi^2}{6} (KT)^2 f'(\mu) +
\frac{7 \pi^4}{360} (KT)^4 f'''(\mu)+...
\end{equation}

\end{document}